\documentclass[journal]{IEEEtran}
\usepackage{graphicx}
\usepackage{float}
\usepackage[caption=false,font=normalsize,labelfont=sf,textfont=sf]{subfig}

\usepackage{amsmath}
\usepackage{amsfonts,amssymb}
\usepackage{dsfont}
\usepackage{algpseudocode}
\usepackage{makecell}
\usepackage{cite}
\usepackage{multirow}
\usepackage{multicol}
\usepackage{booktabs}
\usepackage[ruled,vlined]{algorithm2e}
\usepackage{soul}
\usepackage{tabularx}
\usepackage[table,xcdraw]{xcolor}
\usepackage{rotating}
\usepackage{afterpage}
\usepackage{pdflscape}
\usepackage{url}
\usepackage{titlesec}
\usepackage[english]{babel}
\usepackage[autostyle]{csquotes}
\titleformat{\paragraph}
{\normalfont\normalsize}{\theparagraph}{1em}{}
\titlespacing*{\paragraph}
{0pt}{2ex}{1.5ex}

\titlespacing*{\subsubsection}
{0pt}{2.5ex plus 1ex minus .2ex}{1.5ex plus .2ex}

\usepackage{fancyhdr,lipsum}
\usepackage{hyperref}
\fancypagestyle{mahmood}{%
   \fancyhf{} 
   
   \fancyhead[C]{You can use this material personally. Reprinting or republishing this material for the purpose of advertising or promotion, creating new collective works, reselling or redistributing to servers or lists, or using any copyrighted component in other works must adhere to IEEE policy. The DOI will be supplied as soon as it becomes available.}
}%

\makeatletter
\let\ps@IEEEtitlepagestyle\ps@mahmood
\makeatother

\begin{document}

\title{Discovery of 6G Services and Resources in Edge-Cloud-Continuum}

\author{
    \IEEEauthorblockN{
        Mohammad Farhoudi\textsuperscript{1}, Masoud Shokrnezhad\textsuperscript{1}, Tarik Taleb\textsuperscript{2, 4}, Richard Li\textsuperscript{3}, and JaeSeung Song\textsuperscript{4}\\
    }
    \IEEEauthorblockA{
        \textsuperscript{1} \textit{Oulu University, Oulu, Finland};
        \{mohammad.farhoudi, masoud.shokrnezhad\}@oulu.fi \\
        \textsuperscript{2} \textit{Ruhr University Bochum (RUB), Germany}; tarik.taleb@rub.de
        \\
        \textsuperscript{3} \textit{Southeast University, Nanjing, China}; richard.li@seu.edu.cn \\
        \textsuperscript{4} \textit{Sejong University, Seoul, Korea}; jssong@sejong.ac.kr
    }
}
\maketitle

\begin{abstract}
The advent of 6G networks will present a pivotal juncture in the evolution of telecommunications, marked by the proliferation of devices, dynamic service requests, and the integration of edge and cloud computing. In response to these transformative shifts, this paper proposes a service and resource discovery architecture as part of service provisioning for the future 6G edge-cloud-continuum. Through the architecture's orchestration and platform components, users will have access to services efficiently and on time. Blockchain underpins trust in this inherently trustless environment, while semantic networking dynamically extracts context from service requests, fostering efficient communication and service delivery. A key innovation lies in dynamic overlay zoning, which not only optimizes resource allocation but also endows our architecture with scalability, adaptability, and resilience. Notably, our architecture excels at predictive capabilities, harnessing learning algorithms to anticipate user and service instance behavior, thereby enhancing network responsiveness and preserving service continuity. This comprehensive architecture paves the way for unparalleled resource optimization, latency reduction, and seamless service delivery, positioning it as an instrumental pillar in the unfolding 6G landscape. Simulation results show that our architecture provides near-optimal timely responses that significantly improve the network's potential, offering scalable and efficient service and resource discovery.

\end{abstract}



\section{Introduction}
\label{sec:introduction}

In the wake of the 6G network revolution, ushering in unprecedented advancements in connectivity, a myriad of transformative services emerges,
reshaping people's lives. 
We are entering an era of \textit{ever-increasing device variety and number}, driven by the convergence of Augmented and Virtual Reality (AR/VR) with next-generation wireless networks, robotics, and Artificial Intelligence (AI). Individuals will demand plenty of services, from localized specialized services \cite{6garchitecture_tarik} to immersive metaverse experiences which leads to extreme numbers of Service Requests (SRs). Having real-time devices with differing maximum \textit{data rates} for machine-type communications could exacerbate the problem. Interest in multifaceted applications and services would lead to an ever-growing deluge of \textit{diverse data types} as well. Take, for instance, a user in the metaverse equipped with a Head-Mounted Display (HMD), haptic sensors, and more, simultaneously accessing a diverse array of services -- such as capturing breathtaking vistas with an HMD or obtaining information about their surroundings -- with different types of requests. 

Device and SRs proliferation across various verticals introduces a wide range of \textit{service requirements} \cite{shokrnezhadMetaverse}. Stringent Quality of Service and Quality of Experience (QoS/QoE) standards must be upheld to ensure the prompt delivery of cutting-edge real-time applications.
Anticipating the advent of 6G networks, we envision a landscape characterized by remarkable \textit{dynamism}, marked by frequent changes in user requirements and locations, and the potential for resource or Service Instance (SI) shifts.
Consider our aforementioned user traversing the urban landscapes, dynamically connected to different drones to access their desired services. The perpetual motion of the user and Point of Attachments (PoAs) illustrates future dynamism as a challenge for accessing and using services in real-time. Conceivably, service provisioning is a problem-solution‑fit required to set up and make services available for users. It involves discovering, configuring, deploying, and managing services to ensure they are accessible, functional, and secure. Conventional service provisioning methods fall short when confronted with the impending surge in SRs accompanied by stringent requirements, as their \textit{static} nature fails to adapt swiftly to the fluidity of evolving user needs inherent in the forthcoming 6G network. 

Within the service provisioning framework, Service and Resource Discovery (SaRD) assumes a pivotal role in ever-shifting user bases in a dynamic environment. It encompasses the process of systematically detecting, identifying, registering, and indexing offered network resources, facilitating access to services--tailored to meet specific needs with varied performance, functionality, and availability--in a timely and efficient manner. 
The expansive scope of users, devices, and network coverage demands \textit{scalable} SaRD mechanisms. Considering \textit{automation} and maintaining the \textit{continuity} of discovered services are equally significant, requiring continual adaptation of SaRD processes in response to the dynamic nature. Consequently, traditional SaRD methods are less effective while lacking \textit{intelligence} in discovering the desired SIs. 
In a future network environment where users transmit many requests in various ways, \textit{context-awareness} is crucial. The absence of such awareness hampers the network's ability to determine the relevance of individual requests, contributing to the transmission of excessive information and impeding effective discovery based on real preferences. Hence, the network's ability to discern SR \textit{semantic} from different models becomes essential. Moreover, 
providing access to legitimate services and resources in a \textit{trustless} environment, in which SIs and user trust are unclear, will place an increasing emphasis. 
All in all, there is a pressing demand for a scalable, automated, trustworthy, and efficient foundation for SaRD.

Addressing these vital issues requires the transformation of service provision, especially in SaRD with its upcoming architectures and frameworks, focusing on decentralized methods to enhance scalability. Balanji \textit{et al.} \cite{chordBasedArchitecture} proposed a Service Discovery Management Scheme (SDMS) to enhance availability and reduce SR latencies by clustering services based on functional contexts that adapt to changing consumption patterns. Modern SaRD architectures also leverage semantic information to minimize service search latency.  
Deng \textit{et al.} \cite{bgsd} developed a distributed SaRD using semantic information and graph attention networks to deal with the dynamic environment's service requirements, emphasizing robustness and reducing central cloud reliance. Additionally, some approaches take security-related concerns into account when handling SaRD. Chen \textit{et al.} \cite{blockchainBasedSD} proposed blockchain-based SaRD in the core network that records its function information on a blockchain, ensuring secure registration and discovery while avoiding trust issues.

These studies, while pioneering, fall short of accommodating future futuristic networks' complexities characterized by fluctuating user bases. While some research considered the issues mentioned above through interconnected networks, they often neglect the correlations between these networks and the need for dynamic adaptation, thereby not fully addressing the intertwined issues of scalability and determinism. 
Managing network changes like device additions or removals has been attempted, by locating service or resource directories. Still, future networks require predictive discovery for stringent service requirements, struggling to support dynamicity and maintain service continuity. Current architectures focus on semantics solely for SR matching, overlooking the broader potential of semantic networking, which includes semantic-aware data transmission rather than bit-oriented data and supporting determinism by extracting SR meanings across different modalities.
Finally, existing SaRD architectures do not fully address trustworthy services in large-scale environments, with blockchain's inherent challenges potentially bottlenecking service responses.

This paper bridges the gap by presenting a SaRD architecture for 6G that adeptly navigates the evolving and dynamic environment and paves the way for proactively meeting users' increasing demands. Our contribution consists of the categorization of a SaRD mechanism into distinct parts that lead to adaptive, scalable, intelligent, trustful, and contextually aware discovery. This breakdown fine-tunes service orchestration, showcases platform capabilities, optimizes network infrastructure, and considers application requirements. Our proposed architecture harnesses intelligence by incorporating predictive discovery, anticipating user requests in the next time frame to maintain continuity; semantic networking for precise request understanding--aimed at adapting to user preferences--and efficient transmissions; a robust zoning mechanism ensuring SaRD scalability; and blockchain for decentralized trust across all facets of the discovery process. This comprehensive approach pioneers a resilient 6G SaRD, redefining service provisioning.

The upcoming sections of this paper are arranged coherently as follows. The fundamental principles, inherent challenges, and potential solutions associated with SaRD are explored in Section \ref{sec:background}, and
the proposed SaRD architecture, including its elements, components, and processes, are described in Section \ref{sec:architecture}. In Section \ref{sec:future_directions}, future research directions are presented, and finally, Section \ref{sec:conclusion} offers concluding remarks that encapsulate the study's findings. 

\section{‌Background}\label{sec:background}
 
In the future transformative landscape, and to ensure users' connectivity to enhanced services, 6G will emerge as a beacon of promise, poised to deliver ultra-fast data rates, minimal latency, and extensive connectivity. This vision of ubiquitous connectivity relies on the \textit{edge-cloud-continuum}, revolutionizing service delivery and continuity by integrating cloud and edge computing. This synergy facilitates fluid data movement and processing, which empowers 6G networks to support bandwidth-intensive and real-time applications. Given the integrated infrastructure and the mentioned futuristic 6G edge-cloud-continuum issues,
addressing the multifaceted challenges related to SaRD 
is imperative, which lays a strong foundation for tackling other critical areas like efficiency.



\subsection{Scalability}

Considering the expected number of devices in the near future and their surging demand, SaRD mechanisms' scalability emerges as a key factor. Scalability denotes SaRD's ability to efficiently adapt to the escalating number of users, devices, and SIs within the dynamic edge-cloud-integrated 6G network. A particularly advantageous approach is \textit{Dynamic Overlay Zoning (DOZ)}, also known as sharding or splitting, which dynamically divides networks into distinct zones, each capable of independent operation while interoperating and exchanging information \cite{Hashim2022}. DOZ offers several compelling benefits, making it a desirable choice for dealing with the issue, yet its complexity should be handled carefully. With Machine Learning (ML) algorithms, clustering can be used to predict future demands and adjust the number of zones and their size to proactively enhance network performance and responsiveness, solving the SaRD methods' scalability issue.


\subsection{Determinism}

Determinism, in this context, refers to reliably meeting SRs' needs. In a dynamic and evolving network with user mobility, diverse SRs, and fluctuating resources, the challenge lies in maintaining uninterrupted service flow for desired QoS/QoE. To ensure timely resource access within a futuristic network's dynamic fabric, it is instrumental to access the necessary SRs and plan ahead through \textit{Predictive discovery} \cite{farhoudi2023qosaware}. Anticipating SRs by forecasting them using historical data, traffic patterns, and network feedback elevates the concept of \textit{service continuity} to an innovative and futuristic form. As our Metaverse user transitions between places, they must harness stored supercapacitor energy in smart bricks to meet their energy needs. Predictive discovery forecasts these needs, transmits the SRs along their path, and ensures a continuous, uninterrupted user experience upon arrival.


To achieve precise forecasting and thus the needed service requirements, the SaRD mechanism must interpret \textit{diverse contexts} embedded in requests and manage distinct workloads across various timeframes. Increasingly, users communicate with their surroundings through various means like voice commands, text inputs, and sensory data, requiring an adaptive and automatic method to decipher the SRs' context across varied temporal scopes crucial for streamlined user engagement. \textit{Semantic networking} and communication help extract SR meanings, eliminate superfluous information by exploiting semantic redundancy and ambiguity,
and contribute to determinism by providing a clear understanding of user needs and environmental context. Semantic networking leverages such semantic encoding and decoding techniques as semantic-based calculation and feature extraction to optimize data transmission \cite{a_review_of_general_semanics}. Semantic Service Composition (SeCo) facilitates intelligent decision-making by filtering data and identifying suitable combinations of SIs to meet complex user needs. Specifically, anticipating our Metaverse user's concert request involves predicting SR surges at specific locations through other users' requests with varying modalities.

\subsection{Trustworthiness}

As the network environment evolves towards decentralized and autonomous structures, the challenge lies in upholding trust among various entities. In the context of 6G edge-cloud-continuum, where resource and SI trustworthiness loom as a substantial concern, \textit{blockchain} provides decentralized trust, dynamic orchestration, and security enhancement \cite{fog_based_sd_blockchain}. Using blockchain technology fosters trustless collaboration among decentralized entities that automate cooperative processes, ensuring secure interactions with legitimate services \cite{dynamicSharding}. Our Metaverse user envisions sending SR to a trusted smart brick while mistakenly directing the request to a fake brick could disrupt the experience, leading to disconnection from the concert, highlighting the crucial role of trust, particularly in scenarios like remote surgery.

\section{Proposed Architecture}\label{sec:architecture}
The proposed architecture contributes to service provisioning by providing an efficient SaRD foundation, harnessing 6G's future benefits to offer practical solutions to the issues at hand detailed in this section.
Besides, guidelines for ever-fluctuating SIs and users on how to register their offerings and access their desired services are provided.

\begin{figure}[t!]
    \centering
    \includegraphics[scale=0.09]{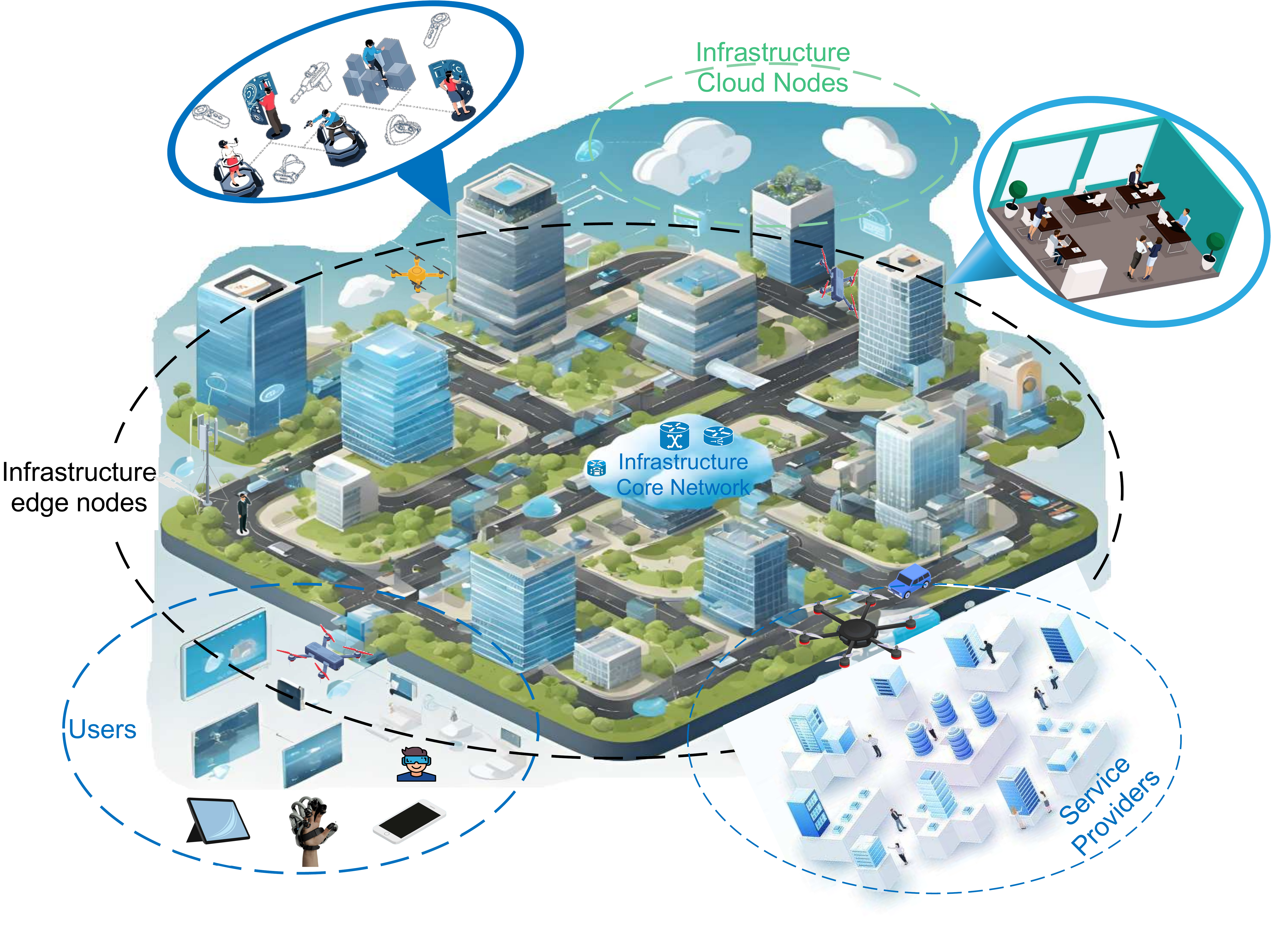}
    \caption{The principal elements or actors shaping the architecture for SaRD in the edge-cloud-integrated 6G network: users, service providers, and infrastructure resource nodes.}
    \label{fig:elements}
\end{figure}

\begin{figure*}[t!]
    \centering
    \includegraphics[scale=0.26]{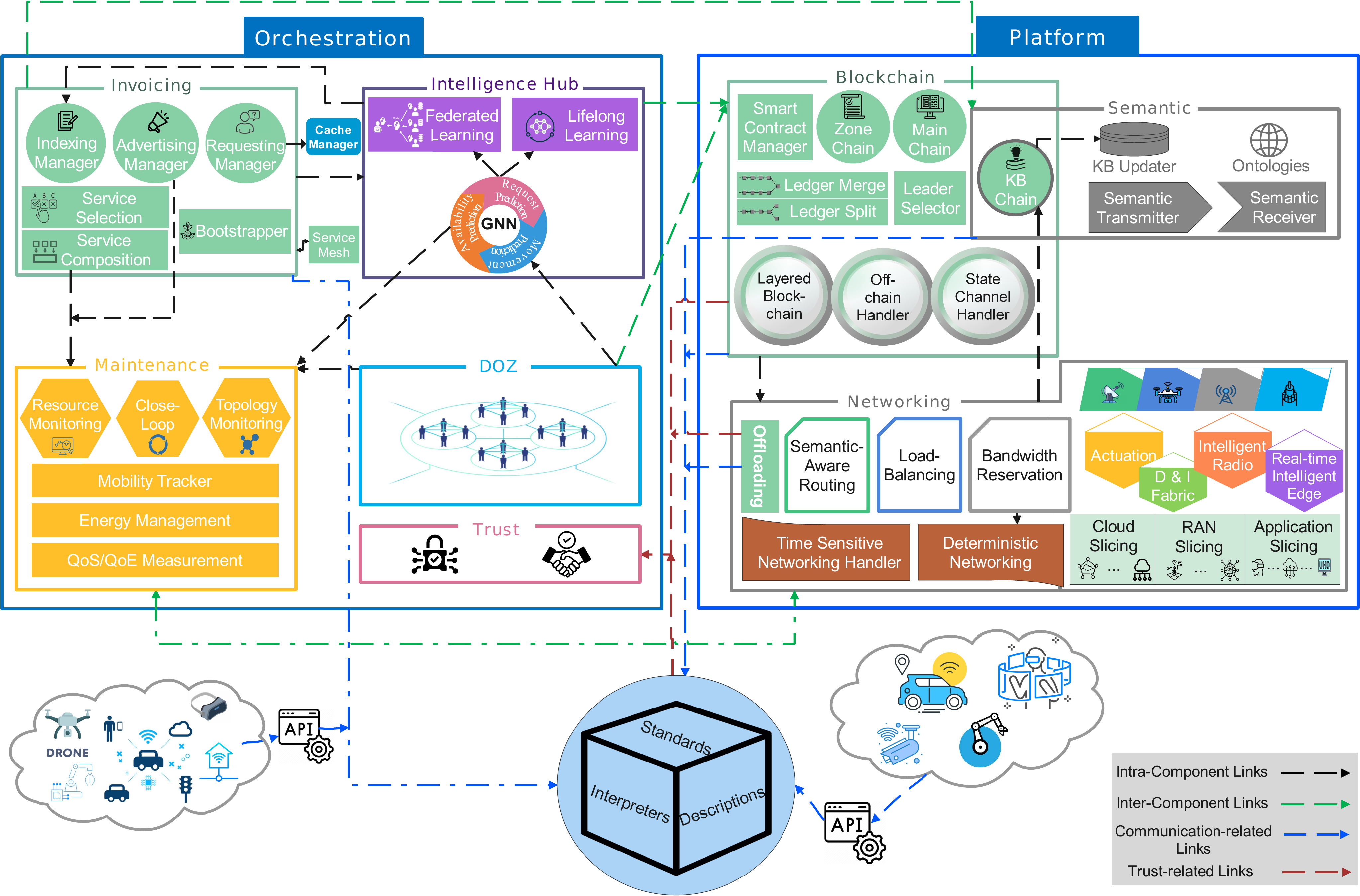}
    \caption{The proposed components of SaRD architecture: essential for achieving scalable, autonomous, trustful, and efficient discovery in the futuristic 6G network.}
    \label{fig:components}
\end{figure*}

\subsection{Elements}
In the rich mosaic 6G SaRD architecture, the \textit{elements} unveil a landscape of actors shaping discovery's functionality, including users, service providers, and infrastructure resource nodes (Fig. \ref{fig:elements}). 
The architecture actors could fluidly transition between different roles. In our Metaverse scenario, for example, a VR headset serves both as a user to request the concert video and as an instance of a service provider to provide real-time data analysis services.

\begin{itemize}
\item \textit{Users} refer to both tangible devices or entities seeking to make use of services. This includes familiar devices like smartphones and extends to immersive technologies like HMDs, neurotech interfaces, and smart contact lenses, all poised to redefine interactions in the Metaverse.

\item \textit{Service Providers} are those elements responsible for providing services. It is up to them to define software components, data models, function chaining maps, and other key parameters based on service description standards. 
They offer their services in physical or virtual--using Virtual Network Functions (VNFs)--instances based on their predefined metrics like cost.
Services can also be composed to offer more complex solutions that aim at fulfilling user wishes that exceed the capabilities of individual atomic services.

\item \textit{Infrastructure Resource Nodes} form the network's foundation, including cloud nodes, extending their reach as infrastructure pillars with networking, storage, computing, and computation capabilities, and edge nodes operate closely to users, enabling low-latency interactions. The continuum includes resource-rich devices like gNodeBs and Metaverse cloud hubs, as well as resource-poor nodes like drones and high-altitude platforms, each contributing distinct capabilities. Network core resources manage critical operations while PoAs connect users to the network. VNF-equipped infrastructure resource nodes aid in networking operations, deploying SIs, and streamlining service provisioning.
\end{itemize}

\subsection{SaRD Components}
The fundamental building blocks of the proposed SaRD architecture dictate how elements contribute to orchestrating services, enabling discovery, access, and delivery of requested services. 
Fig. \ref{fig:components} categorizes components into two parts, each aligned with a specific facet of the SaRD process: Orchestration and Platform.

\subsubsection{Orchestration} \hspace{\parindent} 

The \textit{Orchestration} components are dedicated to decision-making operations. 
Using these components, informed decision processes can be navigated, with scalability, determinism, and trust issues being addressed, adapting to different network conditions, and gaining access to the required information provided by the \textit{Platform} components.



SaRD processes hinge on how various elements interface with the network. Upon network entry, they must acquaint themselves with established standards and essential communication protocols such as learning how to describe services. 
The \textit{Bootstrapper} component launches this process, which enables automatic request submission and interaction through the network, adapting to different device types and user needs. Within SaRD, two request types exist: Advertising Requests (AdRes) and SRs. AdRes are initiated by service providers who seek to promote their services and register their SIs - multiple SIs may offer the same services. Conversely, users issue SRs to access the services they require. To differentiate between these two, SRs are managed by the \textit{Requesting Manager}, whereas AdRes are handled by the \textit{Advertising Manager}. In addition, keeping track of both SRs' and SIs' historical records contributes greatly to discovery processes' intelligent decision-making. To this end, the \textit{Indexing Manager} orchestrates SRs, offered services as well as SIs, service semantic Knowledge Bases (KBs), and resource indexing procedures, utilizing different metrics - such as accuracy, consistency, reliability - to optimize the process. Efficient SaRD operation requires a clear, autonomous interface and a well-functioning indexing manager to maintain real-time knowledge of the network's state.


When SRs are initiated, the process of selecting suitable SIs to fulfill these requests becomes pivotal. The \textit{Service Selection} component, possibly with \textit{Service Composition}, identifies appropriate SI(s) for requests on their path to reach their destinations. The decision requires three sources of information obtained from the \textit{Maintenance} and \textit{Intelligence Hub} components: the roster of available SIs, real-time SI data, and future predictions. Continuous monitoring by \textit{Maintenance} is required to detect SI characteristics and changes such as load fluctuations, mobility, and disconnects, adapting to real-time network conditions. 
For future-oriented predictions, \textit{Intelligence Hub} components predictive analysis and estimation of SIs' availability are leveraged. The decision on which SIs to select is guided by an array of metrics such as the locations of users and SIs, the proximity of SIs to users, the type of service required, current resource utilization levels, energy consumption patterns, and the specific QoS/QoE requirements of the SRs. 


In our SaRD architecture, SI selection aligns with users' requirements, ensuring chosen SIs not only meet but also uphold these standards throughout the service delivery process. In this regard, \textit{QoS/QoE Measurement} component investigates the metrics that affect QoS/QoE through the network, providing a solid foundation for the selection process. User preferences for informed decisions provided by \textit{Maintenance} components can impact SI selection. For instance, the \textit{Energy Management} tracks nodes' power usage (battery and consumption rates) and energy harvesting methods. By allowing our Metaverse users to tap into a variety of energy supply sources based on their preferences, we enabled a knowledgeable and adaptive approach. Moreover, resources and SIs' trustworthiness must be prioritized. The \textit{Trust} component considers the establishment of trust mechanisms using blockchain and defines such trust considerations as the protocols for different entities interacting with smart contracts, the management of these contracts, and the implications of split zoning on trust dynamics. By integrating these considerations, users are empowered to choose SIs that not only adapt to their priorities but are also reliable.

Optimal SI selection and predictive discovery are continuous processes that unfold along the user's path to accessing the required services. The dispersion of users and SIs across the expansive edge-cloud-continuum makes it increasingly effective for segmenting the network into distinct zones to meet the scalability issue. This approach allows users to efficiently access their desired services and connect with appropriate SIs within a specific zone. \textit{DOZ} supports network partitioning while streamlining proper service delivery by grouping requesters and targeting SIs within the same zones. Using network inputs - like real-time feedback, network load, and so forth. - and predictive insights like SI availability predictions, node locations, and SRs prediction, optimal zones are determined. \textit{Intelligence Hub} estimates rezoning times based on user and SI mobility, adapting to dynamic network conditions. DOZ is especially pertinent in scenarios like our Metaverse user experience, where locating an SI globally is impractical. Alternatively, through DOZ, we can investigate the most relevant SIs within the user's zone that significantly contribute to the SaRD approach's scalability.



Several intelligent SaRD processes rely on predictions controlled by \textit{Intelligence Hub} components.
Based on Graph Neural Networks (GNNs) and Deep Reinforcement Learning (DRL) algorithm, \textit{Movement Prediction} anticipates user and SI location predictions, integrating node features such as location, velocity, acceleration, direction, and energy data. Our architectures employ strategies such as as effective sampling, dynamic updates, parallel processing, and hierarchical representations to ensure scalability and handle large-scale graphs in dynamic environments. \textit{Availability Prediction} predicts SI availability in the upcoming timeframe to avoid dispatching SRs to unavailable or soon-to-be-recharging SIs. These components continually adapt to changing user behaviors and external factors through \textit{Continual Learning}, addressing catastrophic forgetting. Further, dynamic zoning coupled with distributed users and SIs emphasizes the need for \textit{Federated Learning}, resulting in shared learning models within these spatial-temporal contexts. All in all, the forward-thinking approach of \textit{Intelligence Hub} components is crucial for maintaining service continuity under dynamic future environments.

\begin{figure}[t!]
    \centering
    \includegraphics[scale=0.41]{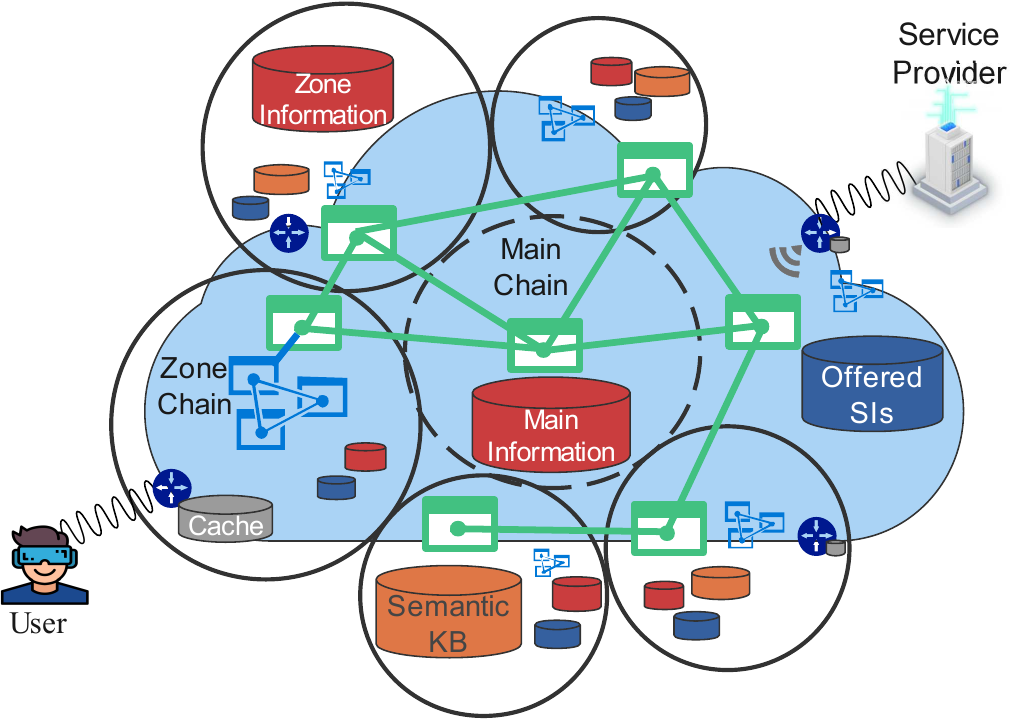}
    \caption{The overall view of four types of indexing crucial for the intelligent SaRD, namely zone state information, request history, offered services and SIs, and service semantic KBs as well as PoAs' caching within the proposed architecture.}
    \label{fig:service_indexing}
\end{figure}

\subsubsection{Platform}  \hspace{\parindent}

Platform components collectively form the underlying foundation required to make informed decisions through the orchestration components. Each of its components provides a dedicated platform that collectively constitutes the bedrock upon which the entire architecture stands and gathers the necessary information for the execution of SaRD functions. 


\begin{figure*}
\captionsetup[subfigure]{font={scriptsize}}
\centering
\subfloat[]{\label{fig:user_flowchart} \includegraphics[width=0.49\textwidth]{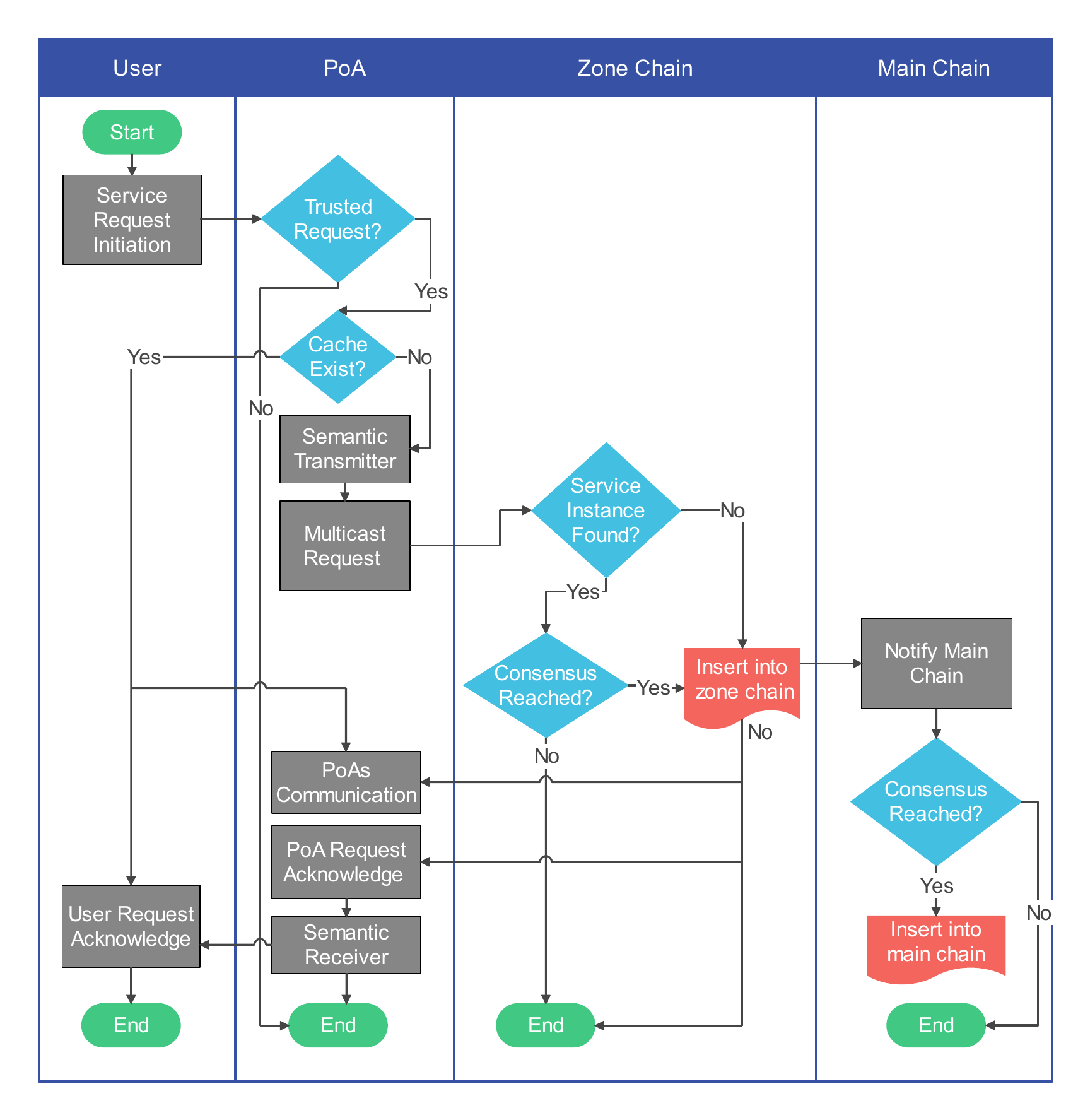}}
\hfill
\subfloat[]{\label{fig:service_flowchart} \includegraphics[width=0.49\textwidth]{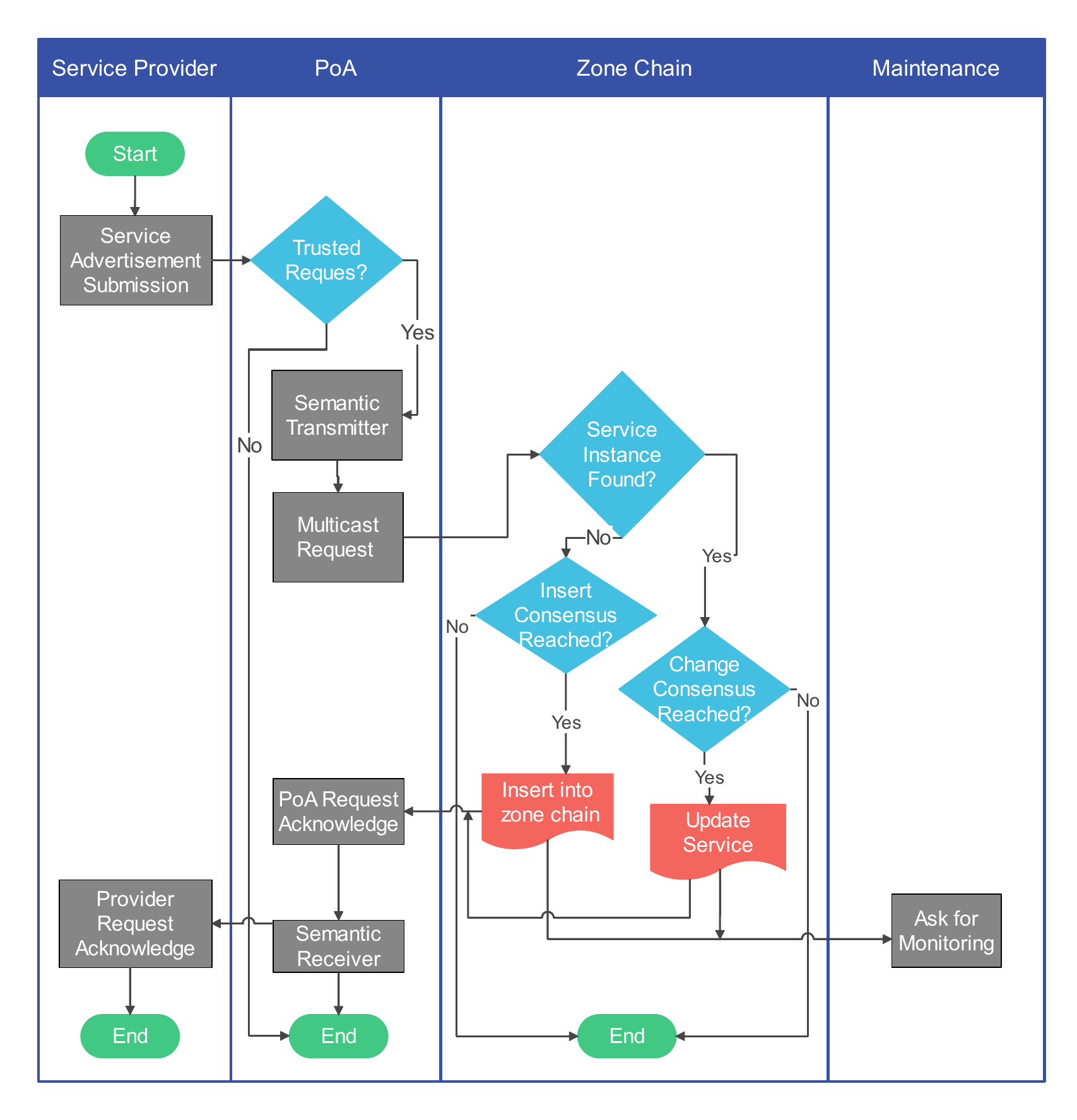}}
\caption{(a) Details the sequence of events in user SRS, starting with SR submission through PoAs. Service instances are selected based on preferences, QoS/QoE, and trustworthiness, with continuity ensured via predictive mechanisms. (b) Outlines the process for service providers to register, modify, or deregister instances. AdRes are submitted through PoAs and instances are added to the zone chain.}
\label{fig:flowchart}
\end{figure*}


To fully enable intelligent decision-making as well as enhance the contextual relevance and efficiency of data transmission in the context of SaRD, \textit{Semantic} components must be incorporated into a holistic platform. The need for semantic awareness arises from the existence of different SRs with varied modalities and fixed meanings, enabling our architecture to handle varied modalities. Hence, merely preserving historical data without semantics proves insufficient in identifying relevant SRs. Semantic communication in our architecture involves the exchange of SR in a meaningful and context-aware manner, where a KB is considered a structured repository that stores semantic information. \textit{Semantic Transmitter} and \textit{Semantic Receiver} components work in tandem to extract knowledge from requests and perform the reverse operation, using semantic fusion methods \cite{semantic_communcation}. The extracted information will be sent as a SR which leads to reduced data transmission as well. The \textit{Semantic} components include detailed KBs, attributes, and semantic annotations, enriching SaRD with accurate, context-aware information. These KBs need to be maintained - within a dedicated \textit{KB Chain} - to facilitate decoding and encoding requests which are continuously updated by \textit{KB Updater} to align with evolving semantic features \cite{shokrnezhad2023semantic}. With up-to-date KBs, Thus, \textit{Semantic} platform contributes significantly to exchanging information and enabling SaRD to its full potential.

As previously highlighted, to enable intelligent decisions in \textit{Intelligence Hub} components, preserving historical SR data and exploring the trustworthiness of users and SIs for informed service selection are crucial. The \textit{Blockchain} components in the architecture ensure trust and data integrity through a multifaceted platform and a hybrid structure combining on-chain and off-chain processing to maintain and manage records efficiently, which is depicted in Fig. \ref{fig:service_indexing}. Each zone has its own \textit{Zone Chain} that retains critical local data such as user states, SI information, attributes, and service semantic KBs that facilitate zone-level SaRD operations. Our architecture employs \textit{Ledger Merge} and \textit{Ledger Split} components to facilitate real-time information exchange and adjustments within \textit{Zone Chains}, ensuring consistent and relevant data after any zone change. On a broader scale, the centralized \textit{Main Chain} offers a high-level, network-wide perspective tasked with summarizing KB statistics; metadata; and filtered, pruned, and aggregated data regarding network states for informed decision-making in DOZ. Alongside, trust is established by deploying blockchain smart contracts, which authenticate each SI and assure users of their reliability, ensuring compliance with established network trust protocols during all interactions and transactions, while dynamically adapting to changes in trust requirements.

Utilizing Blockchain in SaRD introduces challenges such as transaction processing delays and performance bottlenecks due to frequent updates, coupled with inherent latency in transaction handling. To tackle these limitations, our architecture integrates a fast finality mechanism, which accelerates block confirmation; an optimistic roll-up for expedited transaction validation; and state channels while aggregating multiple updates into single transactions. State channels allow direct communication among participants while reducing consensus and blockchain interactions for each update, ensuring adaptability to varying transaction volumes and network conditions.

To establish connections between users and SIs while facilitating SRs and AdRes transmission, a solid networking platform becomes imperative. The incorporation of zoning strategies and semantics introduces novel dimensions to networking, departing from traditional approaches. Semantic-aware routing elevates network operations by introducing an intelligent layer that takes into account contextual information about services and resources when making routing decisions, adapting to dynamic network states. 
SaRD's \textit{Networking} components are imperative in the 6G environment, which uses Network Function Virtualization (NFV) to host virtual SIs, ensuring the delivery of requests to its end with stringent requirements.

\subsection{Discovery Workflows}
The SaRD workflow identifies systematic stages for user, service provider, and infrastructure node engagement in SaRD, outlining step-by-step procedures, communication, and interactions.

As depicted in Fig. \ref{fig:user_flowchart}, the SR process begins with users sending SRs through PoAs. The semantics-related features of the request are extracted by the \textit{Semantic Transmitter} component, operating at the PoA. To determine if the requested service already exists in its cache, the PoA performs a cache check. If found within an acceptable aging window, it furnishes the response to the user. Otherwise, the \textit{Request Prediction} component uses historical data to forecast the user's direction. Accordingly, the PoA propagates the user's intent--extracted features--as a multicast request to the user's targeted location zone. The \textit{Service Selection} component chooses SIs based on the user's preferences, required QoS/QoE, and SI trustworthiness, with prioritized requests being resolved first. The SR, or a bundle of pending SRs, will be added to the SR history on the Zone and Main chains. Simultaneously, the response will be forwarded to the PoA, including the selected SI availability and its transition predictions, driven by the \textit{Availability Prediction} and \textit{Movement Prediction} components. The PoA's \textit{Semantic Receiver} performs the inverse operation of the \textit{Semantic Transmitter}. It promptly delivers responses to the user who initiated the SR, preserving semantic features and relevant data in its cache, subject to aging information. PoAs also communicate with other PoAs for service continuity, forwarding SRs on behalf of users to predicted PoAs along their paths. For SeCo, PoAs share information and send requests for required (atomic) services in advance, ensuring the timely existence of requested instances before users reach their future PoAs.




When a service provider wishes to register, modify, or deregister their instances, they initiate the AdRes process through PoAs using proper service descriptions (Fig. \ref{fig:service_flowchart}). Following the extraction of features, the PoA transmits service information, including instance details. Upon incorporating the offered SI or its modifications into the zone chain, the \textit{Maintenance} components initiate the monitoring process to assess its availability and predictive behavior or cease observation if required. The PoA acknowledges the AdRe's success by confirming it with the service provider.

\section{Performance Evaluation}

\begin{table}[t!]
\caption{Simulation Parameters}
\vspace{-15pt}
\label{table:simulation_paramter}
\begin{center}
\begin{tabular}{|c|c|}
\hline
\textbf{Parameter} & \textbf{Value} \\
\hline
\multirow{2}{*}{Number of infrastructure nodes} & $\sim \mathcal{U}\{5, 25\}$, (connected \\ &  edge-cloud network graph) \\
Number of service providers & $20$ \\
Number of SIs per service provider & $\sim \mathcal{U}\{3, 5\}$ \\
Cost of each link & $\sim \mathcal{U}\{10, 20\}$ \\
Cost of each infrastructure node & $\sim \mathcal{U}\{50, 70\}$ \\
Cost of each SI & $\sim \mathcal{U}\{25, 40\}$ \\
Link capacity (bandwidth) & $\sim \mathcal{U}\{10, 40\}$ Gbps \\
Infrastructure node capacity & $\sim \mathcal{U}\{50, 100\}$ Gbps \\
SI capacity & $\sim \mathcal{U}\{20, 50\}$ Gbps \\
\multirow{4}{*}{User SRs} & User mobility pattern based on \\ & SUMO mobility trace generation \\ & simulator \cite{SUMO2018}, with each user \\ & sending random SRs in each time \\
\hline
\end{tabular}
\end{center}
\vspace{-0.7cm}
\end{table}

\begin{figure*}[t!]\centering
\vspace{0.4cm}
\includegraphics[width=6.1in]{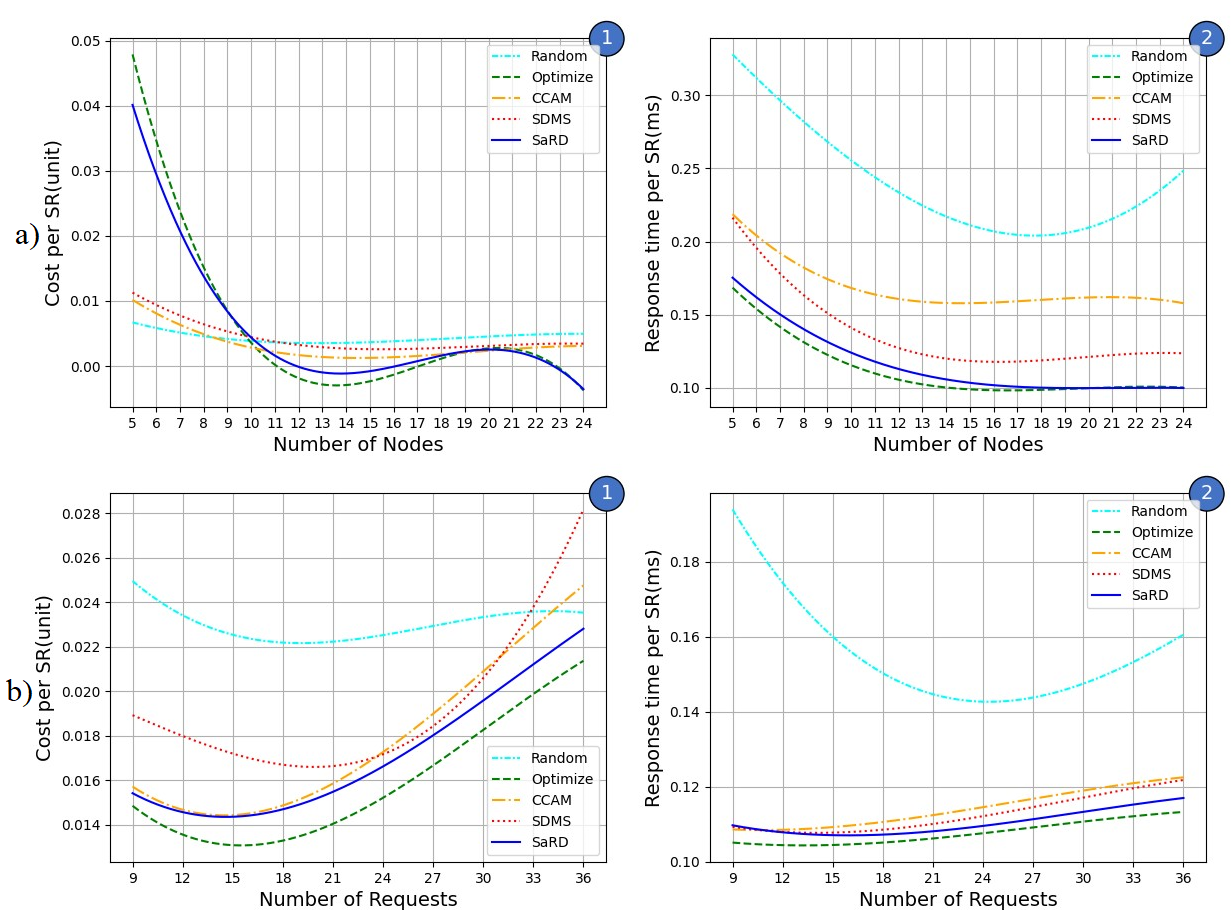}
\vspace{-0.3cm}
  \caption{Comparison of (1) cost and (2) response time for responding to SRs between the proposed SaRD vs. the optimized method, random selection, CCAM \cite{Dang2021}, and SDMS \cite{chordBasedArchitecture} methods, as (a) the number of nodes increases, and (b) the number of requests increases.}
    \vspace{-0.55cm}
    \label{figure:performance}
\end{figure*}


The effectiveness of our SaRD architecture is quantitatively evaluated within a simulated edge-cloud-integrated 6G network considering real and diverse SRs.
The architecture accommodates various PoAs across different zones, mimicking the edge-cloud-continuum, and handles SRs with different modalities leveraging prediction algorithms. SaRD's performance is evaluated with varying numbers of nodes and SRs, where not only do individuals frequently change locations but also their data rates are varyingly adjusted to reflect the required dynamism. The evaluation assesses key metrics such as node, SI, and link costs, as well as SRs response time, providing a detailed understanding of our architecture's performance in a dynamic setting.
Simulation parameters, described in Table \ref{table:simulation_paramter}, offer flexibility to accommodate architectural elements.














The proposed architecture partitions the network based on real-time data and prediction info, ensuring demand fluctuations do not adversely affect the architecture. Blockchain safeguards SIs' trustworthiness with main and zone chains during the selection process. Additionally, semantic encoding/decoding is explored in the PoAs in which extracted features of requests are utilized as transmitted requests - beyond their modalities - for discovery using \textit{KB chain}, enhancing prediction accuracy by tracking relevant SR history. Our implemented DRL algorithms continually predict each node's availability and each PoA's SRs, optimizing resource allocation to meet stringent QoS/QoE requirements and maintain service continuity in the dynamic environment.


Our architecture is benchmarked against alternative SaRD methods, including optimal SI and node assignments, random selection, the approach in \cite{Dang2021} for Connected and Cooperative Autonomous Mobility (CCAM), and SDMS for IoT ad hoc networks \cite{chordBasedArchitecture}. Results in Fig. \ref{figure:performance} highlight response time and cost as performance metrics. While the optimal method excels at ideal SR assignment due to its omniscient knowledge of actual requests, its complexity--considering all combinations of SRs and resource allocations on different nodes--makes it impractical for timely solutions. Random selection is simple, yet results in high costs and a high rate of unsupported requests, leading to poor response times. The CCAM method performs better in terms of cost in networks with few nodes because of their inability to support all requests effectively, especially as the number of SRs and nodes increases. Unlike CCAM, which struggles with scalability and cannot serve single services on a single node, and SDMS, which focuses on service availability and overlooks allocation costs, our method, with its simplicity, scalability, and context-awareness, achieves significant efficiency gains by predicting future SRs, resulting in better allocation and enhanced response times. SR continuity is achieved by intelligently accommodating fluctuating SR volumes over time, providing the adaptability needed in a dynamic environment. Besides, our approach's zoning methodology demonstrates necessary scalability, maintaining high SR acceptance rates even as request numbers and data rates rise. This resilience is notable given the dynamic nature of SIs and users, including their mobility and subsequent zone transitions. 

\section{Future Research Directions}\label{sec:future_directions}

The proposed research directions for the future of SaRD focus on three key domains: semantic-aware; scalable; and predictive discovery, each offering unique opportunities to shape the future landscape.

Future research should examine semantic-aware SaRD's efficiency and performance. Ensuring semantic data sharing privacy, optimizing semantic fusion, and investigating energy-efficient methods for semantic processing are vital. In the context of a dynamic semantic environment, synchronization, self-healing, and semantic reconciliation techniques become crucial to be explored. 
Further considerations could include energy-saving methods for semantic training and using resource-rich nodes for this step. 

To enhance the scalability and service continuity of SaRD, future research should delve into DOZ and address blockchain-related challenges. Exploring interoperable smart contracts and employing strategies for efficient ledger updates following zone changes, as discussed in \cite{fahim2023blockchain}, could minimize zoning intervals, which would ultimately translate to improved service continuity. Also, dynamically adjusting chain block size could prevent a performance bottleneck. It would be a priority to optimize off-chain operations, considering overhead and utility. Additionally, a balance between maintaining historical chain data for accurate prediction and addressing storage concerns must be struck. 
Lastly, in a trustless environment, preventing unauthorized access and privilege escalation during bootstrapping becomes paramount for ensuring the integrity of the architecture.

For the goal of more suitable predictive discovery, the challenges posed by federated learning, such as heterogeneity of data, domain shifts, model stability, and the interplay between continual and federated learning must be addressed. In federated learning, privacy-preserving methods should be of paramount importance, safeguarding exchange models during data exchange. The incorporation of continual and federated learning with GNNs introduces complexities related to model stability, optimization, and complexity that require investigation to enhance prediction accuracy. Meta-learning techniques should also be explored to improve SaRD learning capabilities. 
\section{Conclusion}
\label{sec:conclusion}
In a world on the brink of the 6G revolution, the service and resource discovery architecture proposed herein emerges as a linchpin for future service provisioning in a highly dynamic environment. As a result of the architecture's orchestration, blockchain-backed trust, semantic networking, and pioneering predictive capabilities, powered by graph neural networks and deep reinforcement learning, it is positioned at the forefront of innovation in the discovery domain. The blockchain infrastructure, encompassing different zone chains and a main chain, delivers an immutable record that aids the learning algorithm in performing intelligent predictive discovery while accommodating dynamic zone changes. Crucially, the incorporation of dynamic overlay zoning imparts not only resource optimization and latency reduction but also scalability, adaptability, and resilience—essentials for a future defined by evolving service demands. Our learning algorithms predict user and service instance movement patterns and anticipate service requests using continual learning and privacy-preserving federated learning across different zones to cope with futuristic dynamicity. Beyond its immediate applications, it lays the groundwork for a more responsive, efficient, and reliable 6G edge-cloud-continuum that preserves service continuity in the future.


\section*{Acknowledgment}
Prof. Tarik Taleb and JaeSeung Song are co-corresponding
authors of this work. This work was partially conducted at ICTFICIAL Oy. It is partially supported by the European Union’s HE research and innovation program HORIZON-JUSNS-2023 under the 6G-Path project (Grant No.
101139172). It is also partially supported by the Business Finland 6Bridge 6Core project under Grant No. 8410/31/2022. The paper reflects only the authors’ views, and the European Commission bears no responsibility for any utilization of the information contained herein.


\bibliographystyle{unsrt}
\bibliography{bibliography.bib}

\begin{IEEEbiographynophoto}{Mohammad Farhoudi} (Student Member, IEEE) received his Master's degree in Computer Networks from the Amirkabir University of Technology, Tehran, Iran, in 2017. He is currently pursuing a Ph.D. at the Centre for Wireless Communications, University of Oulu, Finland. He has served as a Senior Research Assistant at the Amirkabir University of Technology, working on a national project focused on utilizing SDN and NFV in large-scale data and transport networks. Additionally, he has over eight years of professional experience with Kilid Gostar Bina, Saba System Sadra, and Fava Pars Company, specializing in computational networking and computer security. His research interests include semantic-aware service orchestration, edge-cloud networks, and AI/ML mechanisms for service discovery and resource allocation.
\end{IEEEbiographynophoto}

\begin{IEEEbiographynophoto}{Masoud Shokrnezhad}
received his Ph.D. degree (recognized as a bright talent) in computer networks from Amirkabir University of Technology (Tehran Polytechnic), Tehran, Iran, in 2019. He is currently a postdoctoral researcher with the Center of Wireless Communications at the University of Oulu, Finland. Between June 2021 and December 2021, he was a postdoctoral researcher at the School of Electrical Engineering, Aalto University, Espoo, Finland. Prior to that, he worked as a senior system designer and engineer with FavaPars and Pouya Cloud Technology in Tehran, Iran, since 2013. Throughout his career, Dr. Shokrnezhad has been involved in numerous national, international, and European projects focused on designing and developing computing and networking frameworks. He has also co-managed a startup that develops SDWAN solutions for B2B use cases.
\end{IEEEbiographynophoto}

\begin{IEEEbiographynophoto}{Tarik Taleb}
is currently a Full Professor at Ruhr University Bochum, Germany. He was a Professor with the Center of Wireless Communications (CWC), University of Oulu, Oulu, Finland. He is the founder of ICTFICIAL Oy, and the founder and the Director of the MOSA!C Lab, Espoo, Finland. From October 2014 to December 2021, he was an Associate Professor with the School of Electrical Engineering, Aalto University, Espoo, Finland. Prior to that, he was working as a Senior Researcher and a 3GPP Standards Expert with NEC Europe Ltd., Heidelberg, Germany. Before joining NEC and till March 2009, he worked as an Assistant Professor with the Graduate School of Information Sciences, Tohoku University, in a lab fully funded by KDDI. From 2005 to 2006, he was a Research Fellow with the Intelligent Cosmos Research Institute, Sendai. He received the B.E. degree (with distinction) in information engineering and the M.Sc. and Ph.D. degrees in information sciences from Tohoku University, Sendai, Japan, in 2001, 2003, and 2005, respectively.
\end{IEEEbiographynophoto}

\begin{IEEEbiographynophoto}{Richard Li}
is Chair Professor of Network Technologies, School of Computer Science and Engineering, Southeast University, Nanjing, China. Prior to joining Southeast University in April 2024, Richard worked with Futurewei Technologies, aka Huawei R\&D USA, from 2007 to 2024 as Chief Scientist, SVP, and Head of Network Technologies Lab in the San Francisco Bay Area, USA. Before that, Richard worked with Cisco and Ericsson on their networking products, technologies, solutions and network operating systems. Richard also served as the Chairman of the ITU-T FG Network 2030 from 2018 to 2020, the Vice Chairman of the Europe ETSI ISG Next-Generation Protocols from 2016 to 2019, and Chairs of steering committees and technical program committees of some academic and industrial conferences. Currently, he serves on the advisory board for IEEE IoT Journal, as a technical editor for IEEE Network Magazine, and as guest editors for special issues of some journals and magazines.
\end{IEEEbiographynophoto}

\begin{IEEEbiographynophoto}{Jaesung Song}
is a full professor in the Department of Computer \& Information Security Sejong University. He holds the position of Technical Plenary Vice Chair of the oneM2M global IoT standards initiative. Prior to his current position, he worked for NEC Europe Ltd. and LG Electronics in various positions. He received a Ph.D. at Imperial College London in the Department of Computing, United Kingdom. He holds B.S. and M.S. degrees in computer science from Sogang University. His research interests span the areas of beyond 5G and 6G, AI/ML enabled network systems, software engineering, networked systems and security, with focus on the design and engineering of reliable and intelligent IoT/M2M platforms.
\end{IEEEbiographynophoto}

\end{document}